\documentstyle[prl,aps,epsfig,amsmath,amssymb,multicol]{revtex}

\newcommand{\eq}{\!=\!}
\newcommand{\defeq}{\stackrel{\text{def}}{=}}
\newcommand{\quadeq}{\quad\phantom{{}={}}}
\renewcommand{\phi}{\varphi}
\renewcommand{\L}{\Lambda}
\renewcommand{\l}{\lambda}
\newcommand{\B}{{\mathcal B}}

\newcommand{\bra}[1]{\langle#1|}
\newcommand{\ket}[1]{|#1\rangle}
\newcommand{\Ppsi}{P_{\ket{\psi}}}
\newcommand{\Pdelta}{P_{\ket{\delta}}}
\newcommand{\sw}[1]{#1\rangle\langle#1}
\newcommand{\pfbox}{\hfill$\blacksquare$\qquad}

\begin{document}
\title{Strict detector-efficiency bounds for $n$-site Clauser-Horne
  inequalities}
 
\author{Jan-{\AA}ke Larsson$^{1,2,*}$ and Jason Semitecolos $^{2
    ,\dagger } $ } \address{$^1$Matematiska Institutionen,
  Link\"opings Universitet, SE-581 83 Link\"oping, Sweden}
\address{$^2$Centre for Quantum Computation, Clarendon Laboratory,
  University of Oxford, Parks Road, Oxford OX1 3PU, UK}

\date{\today}
\maketitle
\begin{abstract}
  An analysis of detector-efficiency in many-site Clauser-Horne
  inequalities is presented, for the case of perfect visibility. It is
  shown that there is a violation of the presented $n$-site
  Clauser-Horne inequalities if and only if the efficiency is greater
  than $\frac{n}{2n-1}$. Thus, for a two-site two-setting experiment
  there are no quantum-mechanical predictions that violate local
  realism unless the efficiency is greater than $\frac{2}{3}$.
  Secondly, there are $n$-site experiments for which the
  quantum-mechanical predictions violate local realism whenever the
  efficiency exceeds $\frac{1}{2}$.
\end{abstract}

\begin{figure}[htbp] 
  \begin{center}
    \psfig{file=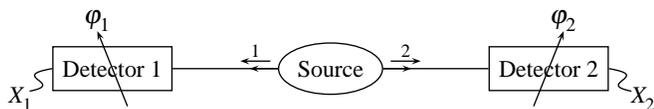}
    \caption{A schematic picture of the physical system used in the
      two-site case. The source emits entangled pairs of particles,
      one at each detector site. The detectors have a local (angular)
      setting $\phi_i\in\{a,\,b\}$.}
    \label{fig:system}
  \end{center}
\end{figure}

\begin{multicols}{2}
The Bell, Clauser-Horne-Shimony-Holt (CHSH), and Clauser-Horne (CH)
\cite{BELL} inequalities state what correlations or probabilities are
to be expected from a local realistic model.  The experimental system
used to test these inequalities (or rather, their prerequisites) is
shown schematically in figure~\ref{fig:system}.  Tests made to date
are subject to different loopholes such as reduced efficiency, reduced
visibility, and certain problems of obtaining strict locality, making
the application of the above inequalities depend on extra assumptions.
The search for experiments where less assumptions are needed have
reached quite far, but one loophole remains: detector inefficiency,
which is usually dealt with by making the no-enhancement assumption.
The present paper is an analysis without use of this assumption of
efficiency bounds for the CH inequality and some $n$-site
generalizations of it.

The formalism of probability theory will be used below, where the
event space $\L$ is a space of points $\l$ corresponding to a certain
value of the ``hidden variable''. The measurement results are
described by random variables which take their values in the value
space $V$ (here $\{0,\,1\}$), and on the space there is a probability
measure $P$, which gives the probabilities of events, i.e., subsets of
$\Lambda$.

The Bell and CHSH inequalities are relations on correlations, and
contain various restrictions on the measurement results. For example,
in the case of the Bell inequality, perfect correlation at equal
detector settings and the labeling $\pm1$ of the results, which makes
it unsuitable for use in experiments where noise and inefficiency
occurs. The CHSH inequality may be used in noisy situations, but still
cannot directly accommodate inefficiencies of the measurement setup
(this can be remedied; see \cite{EFFICIENCY}). The CH inequality, on
the other hand, is a relation on probabilities and does not need such
restrictions on the results \cite{EVENTS} but only the prerequisites
of Realism and Locality.

\emph{Theorem 1: (Clauser-Horne)} If we have \label{thm:CH}
\begin{enumerate}
\item \label{Bp1} \emph{Realism.} Measurement results can be
  described by probability theory, using (two families of) random
  variables.
  \begin{equation*}
    \begin{split}
      X_1(\phi_1,\phi_2):\L&\rightarrow V\\
      \l&\mapsto X_1(\phi_1,\phi_2,\l)\\
      X_2(\phi_1,\phi_2):\L&\rightarrow V\\
      \l&\mapsto X_2(\phi_1,\phi_2,\l)
    \end{split}
    \qquad\quad \forall \phi_1,\phi_2.
  \end{equation*}
\item \label{Bp2} \emph{Locality.} A measurement result at one site
  should be independent of the detector orientation at the other
  site,
  \begin{equation*}
    \begin{split}
      X_1(\phi_1,\l)&\defeq X_1(\phi_1,\phi_2,\l),\text{
        independently of }\phi_2\\
      X_2(\phi_2,\l)&\defeq X_2(\phi_1,\phi_2,\l),\text{
        independently of }\phi_1\\
    \end{split}
  \end{equation*}
\end{enumerate}
except at a null set, then, with $A_i=X_i(a,\l)$ and
$B_i=X_i(b,\l)$,
\begin{multline}
  P(A_1\eq B_2\eq 1)+P(B_1\eq A_2 \eq 1)
  -P(B_1\eq B_2 \eq 1)\\
  \le P(A_1\eq 1)+P(A_2 \eq 1) -P(A_1\eq A_2\eq 1)
\end{multline}

\emph{Proof:} Given (i), the following inequality is obviously true:
\begin{multline}
  P(A_1\eq B_2\eq 1 \cup B_1\eq A_2 \eq 1)
  \le  P(A_1\eq 1 \cup A_2 \eq 1).\label{eq:obv2}
\end{multline}
On the right-hand side we have
\begin{equation}
  \begin{split}
    &P(A_1\eq 1 \cup A_2 \eq 1)\\
    &\quad=P(A_1\eq 1)+P(A_2 \eq 1) -P(A_1\eq A_2\eq 1),
  \end{split}
\end{equation}
and on the left-hand side, by use of (ii),
\begin{equation}
  \begin{split}
    &P(A_1\eq B_2\eq 1 \cup B_1\eq A_2 \eq 1)\\
    &\quad= P(A_1\eq B_2\eq 1)+P(B_1\eq A_2 \eq 1)\\
    &\quadeq-P(A_1\eq B_1\eq A_2\eq B_2 \eq 1)\\
    &\quad\ge P(A_1\eq B_2\eq 1)+P(B_1\eq A_2 \eq 1)\\
    &\quadeq-P( B_1\eq B_2 \eq 1).\hspace{2.5cm}\blacksquare
  \end{split}
\end{equation}

This inequality is well suited for direct inclusion of noise and
inefficiency, as can be seen in \cite{EBERHARD}. There, the
Clauser-Horne inequality given in a slightly different form is used to
investigate if it is violated by quantum mechanics at lower
efficiencies than the $82.83\%$ bound obtained from the CHSH
inequality \cite{EFFICIENCY}, and at which level of background noise
this is possible.  The calculation in \cite{EBERHARD} uses the
quantum-mechanical expressions for the (ideal) probabilities, e.g.
\begin{gather}
  \stepcounter{equation} \Ppsi (A_1\eq A_2\eq 1)
  =\bra{\psi}\sw{1_{A_1}1_{A_2}}\ket{\psi},
  \label{eq:P2}\tag{\theequation a}\\
  \Ppsi (A_1\eq 1)=\bra{\psi} \left(|\sw{1_{A_1}}|\otimes{\Bbb
      I}_2\right) \ket{\psi},\label{eq:P1}\tag{\theequation b}
\end{gather}
to derive a quantum measurement operator ${\mathcal B}$ corresponding
to the probabilities in the CH inequality including detector
inefficiency:
\begin{eqnarray}
  \bra{\psi}{\mathcal B}\ket{\psi}\notag
  &=&\eta^2\Big(\Ppsi(A_1\eq A_2\eq 1)  
  +\Ppsi(A_1\eq B_2\eq 1)\notag\\
  &&\phantom{\eta^2\Big(}+\Ppsi(B_1\eq A_2 \eq 1)
  -\Ppsi(B_1\eq B_2 \eq 1)\Big)\notag\\
  &&- \eta\Big(\Ppsi(A_1\eq 1)+\Ppsi(A_2 \eq 1)\Big).
  \label{eq:B}
\end{eqnarray}
Note that the assumption of independent errors at a constant rate is
used in equation~(\ref{eq:B}), and this assumption will be used from
here on in this paper. In \cite{EBERHARD}, the parameters of
${\mathcal B}$ (efficiency $\eta$ and detector settings $\phi_i$) and
$\ket{\psi}$ are varied randomly to show that there is a violation at
no background if $\eta> \frac{2}{3}$.

To obtain this bound analytically, the eigenvalues of ${\mathcal B}$
would be needed, and the calculation is manageable.  Unfortunately,
this involves solving a fourth-degree polynomial equation, and since
the degree of the polynomial will increase rapidly with $n$, a method
more suitable for the purpose of this paper will be presented. The
two-site bound $\frac{2}{3}$ will be obtained analytically without too
much calculation, and this will be generalized below.

\emph{Theorem 2: (The lowest possible efficiency bound from the CH
  inequality)} In the case of independent errors at a constant rate,
there is a violation of the CH inequality if and only
if\label{thm:2CHbound}
\begin{equation*}
  \eta>\frac{2}{3}.
\end{equation*}

\emph{Proof:} An important observation is that the CH inequality with
the quantum probabilities from (\ref{eq:B}) inserted is equivalent to
\end{multicols}
\noindent\hspace{.1\textwidth}%
\makebox[.4\textwidth]{\hrulefill}\vrule\rule{0pt}{2ex}
\begin{eqnarray}
  \eta\le\frac{\Ppsi(A_1\eq 1)+\Ppsi(A_2 \eq 1)}{\Ppsi(A_1\eq A_2\eq 1)  
    +\Ppsi(A_1\eq B_2\eq 1)
    +\Ppsi(B_1\eq A_2 \eq 1)
    -\Ppsi(B_1\eq B_2 \eq 1)}.\label{eq:2bound}
\end{eqnarray}
\vspace{-1ex} 

\noindent
\hspace{.5\textwidth}
\vrule\rule{0pt}{2ex}%
\raisebox{2ex}{\makebox[.4\textwidth]{\hrulefill}}
\begin{multicols}{2}
Clearly,
\begin{align}
  \stepcounter{equation}
  & \Ppsi( A_1\eq A_2\eq 1)\le \min_{i=1,2} \Ppsi(A_i \eq 1)
  \label{eq:proba}\tag{\theequation a}\\
  & \Ppsi( A_1\eq B_2\eq 1)\le \Ppsi(A_1 \eq 1)\tag{\theequation b}\\
  & \Ppsi( B_1\eq A_2\eq 1)\le \Ppsi(A_2 \eq 1)\tag{\theequation c}\\
  & \Ppsi(B_1\eq B_2\eq 1)\ge 0.\label{eq:probd}\tag{\theequation d}
\end{align}
The lowest possible bound in (\ref{eq:2bound}) would be obtained
when we have equality in (\ref{eq:proba}--d), and when $\Ppsi(A_1 \eq
1)=\Ppsi(A_2 \eq 1)$ (which gives the best possible value in
(\ref{eq:proba}) \cite{EQUAL}).  We then would have
\begin{equation}
  \eta\le\frac{2\Ppsi(A_1\eq 1)}{3\Ppsi(A_1\eq 1)}=\frac{2}{3}.
\end{equation}
A lower bound cannot be obtained; we have proved the \emph{only if}
part. 

It is not possible to devise a quantum state giving equality in
(\ref{eq:proba}--d) that violates the CH inequality \cite{SURPRISE},
but there are states that come arbitrarily close, and such states
will be used in the proof of the \emph{if} part.  Given $\epsilon>0$
and using $\theta=2\arctan(\epsilon)$, the quantum state (see
\cite{PROOF}),
\begin{eqnarray}
  \ket{\delta}&=&C\Big(\big(1-2\cos(\theta)\big)\ket{0_{B_1}0_{B_2}}
  \notag\\
  &&\qquad+\sin(\theta)\big(\ket{1_{B_1}0_{B_2}}
  +\ket{0_{B_1}1_{B_2}}\big)\Big)
\end{eqnarray}
and the rotation
\begin{equation}
  \begin{pmatrix}
    \ket{0_{A_i}}\\\ket{1_{A_i}}
  \end{pmatrix}
  =
  \begin{bmatrix}
    \cos(\theta)&-\sin(\theta)\\\sin(\theta)&\cos(\theta)
  \end{bmatrix}
  \begin{pmatrix}
    \ket{0_{B_i}}\\\ket{1_{B_i}}
  \end{pmatrix}
\end{equation}
yields
\begin{align}
  \stepcounter{equation}
  &\Pdelta( A_1\eq A_2\eq 1)=K\neq0
  \label{eq:K2a}\tag{\theequation a}\\
  &\Pdelta( A_1\eq B_2\eq 1)=K
  \tag{\theequation b}\\
  &\Pdelta( B_1\eq A_2\eq 1)=K
  \tag{\theequation c}\\
  & \Pdelta(B_1\eq B_2\eq 1)=0
  \tag{\theequation d}\\
  &\Pdelta(A_1\eq 1)=\Pdelta(A_2 \eq 1)=K(1+\epsilon^2),
  \tag{\theequation e}
\end{align}
that is, (\ref{eq:2bound}) reduces to
\begin{equation}
  \eta\le\frac{2K(1+\epsilon^2)}{3K}=\frac23(1+\epsilon^2).
\end{equation}
So when $\eta>\frac{2}{3}$, there exists a quantum state that will
give a violation of the CH inequality, which proves the \emph{if}
part.\pfbox

A note here is that the $\ket{\delta}$ used above is \emph{not} an
eigenvector to $\B$ as used in \cite{EBERHARD}. A comparison shows
that using the best possible $\ket{\delta}$ at a given $\eta$ gives a
violation about 80\% of the violation from the eigenvector
\cite{HWANG}.

In the many-site case, a number of extensions of the CH inequality are
possible. In the two-site case above, the inequality uses one- and
two-detection probabilities. To generalize, a choice is made yielding
a simple expression that contains only $(n-1)$- and $n$-detection
probabilities. The sums used below denote a summation over all
possible combinations.

\emph{Theorem 3: (An $n$-site Clauser-Horne inequality)} If we have
Theorem~1 (\ref{Bp1})--(\ref{Bp2}) except at a null set, then
\begin{multline}
  \sum P(\text{One }B\text{ and all other }A\text{'s}\eq 1)\\
  -\sum P(\text{Even number of }B\text{'s and all
      other }A\text{'s}\eq 1 )\\
  \le\sum P(\text{All }A\text{'s}\eq 1\text{ except one})
  -(n-1) P(\text{All }A\text{'s}\eq 1)
\end{multline}

\emph{Proof:} The following inequality is obviously true:
\begin{multline}
  P\big(\cup\{\text{One }B\text{ and 
      all other }A\text{'s}\eq 1\}\big)\\
  \le P\big(\cup\{\text{All }A\text{'s}\eq 1\text{ except 
      one}\}\big)
\end{multline}
On the right-hand side we have 
\begin{multline}
  P\big(\cup\{\text{All }A\text{'s}\eq 1\text{ except one}\}\big)\\
  = \sum P(\text{All }A\text{'s}\eq 1\text{ except one})
  - (n-1)P(\text{All }A\text{'s}\eq 1),
\end{multline}
and on the left-hand side,
\begin{eqnarray}
  &&P\big(\cup\{\text{One }B\text{ and
        all other }A\text{'s}\eq 1\}\big)\notag\\
    &&\quad= \sum P(\text{One }B\text{ and
        all other }A\text{'s}\eq 1)\notag\\
    &&\quadeq- \sum P(\text{Two }B\text{'s and
        \emph{all} }A\text{'s}\eq 1)\notag\\
    &&\quadeq+ \sum P(\text{Three }B\text{'s and
        \emph{all} }A\text{'s}\eq 1)\notag\\
    &&\quadeq-\quad \ldots\notag\\
    &&\quadeq +(-1)^n P(\text{All }B\text{'s and \emph{all}
        }A\text{'s}\eq 1).
\end{eqnarray}
Now
\begin{multline}
  0 \le P(\text{Some set of }B\text{'s and \emph{all}
      }A\text{'s}\eq 1) \\
  \le P(\text{The same }B\text{'s and all other }A\text{'s}\eq 1),
\end{multline}
and therefore, taking the signs into account
\begin{eqnarray}
  && P\big(\cup\{\text{One }B\text{ and
    all other }A\text{'s}\eq 1\}\big)\notag\\
  &&\quad\ge \sum P(\text{One }B\text{ and
    all other }A\text{'s}\eq 1)\notag\\
  &&\quadeq- \sum P(\text{Two }B\text{'s and
    all other }A\text{'s}\eq 1)\notag\\
  &&\quadeq- \sum P(\text{Four }B\text{'s and
    all other }A\text{'s}\eq 1)\notag\\
  &&\quadeq- \quad\ldots\notag\\
  &&\quadeq- 
  \begin{cases} 
    P(\text{All }B\text{'s}\eq 1),& n\text{ even,}\\
    \sum P(\text{One }A\text{ and all other }B\text{'s}\eq
    1), & n\text{ odd.}
  \end{cases}
\end{eqnarray}
which concludes the proof.\pfbox

In the same spirit as in the two-site case (Theorem~2), we have

\emph{Theorem 4: (The lowest possible efficiency bound from the above
  $n$-site CH inequality)} In the case of independent errors at a
constant rate, there is a violation of the above $n$-site CH
inequality if and only if\label{thm:nCHbound}
\begin{equation*}
  \eta>\frac{n}{2n-1}.
\end{equation*}

\emph{Proof:} Again, the $n$-site CH inequality above with quantum
probabilities inserted is equivalent to 
\end{multicols}
\noindent\hspace{.1\textwidth}%
\makebox[.4\textwidth]{\hrulefill}\vrule\rule{0pt}{2ex}
\begin{equation}
  \eta\le\frac{\sum\Ppsi(\text{All }A\text{'s}
    \eq 1\text{ except one})}{
    \sum\Ppsi(\text{One }B
    \text{ and all other }A\text{'s}\eq 1)
    +(n-1) \Ppsi(\text{All }A\text{'s}\eq 1)
    -\sum \Ppsi(\text{Even number of }B\text{'s and all
      other }A\text{'s}\eq 1)}.\label{eq:nbound}
\end{equation}
\vspace{-2.5ex} 

\noindent
\hspace{.5\textwidth}
\vrule\rule{0pt}{2ex}%
\raisebox{2ex}{\makebox[.4\textwidth]{\hrulefill}}
\begin{multicols}{2}
Clearly,
\begin{align}
  \stepcounter{equation}
  & \Ppsi(\text{All }A\text{'s}\eq 1)
  \le \min \Ppsi(\text{All }A\text{'s but one} \eq 1)
  \label{eq:nproba}\tag{\theequation a}\\
  & \Ppsi(\text{One }B
  \text{ and all other }A\text{'s}\eq 1)\notag\\
  &\qquad\qquad\qquad\le\Ppsi(\text{The same }A\text{'s} \eq 1)
  \tag{\theequation b}\\
  & \Ppsi(\text{Two or more }B\text{'s}\eq 1)\ge 0,
  \tag{\theequation c}
\end{align}
The lowest possible bound in (\ref{eq:nbound}) would be obtained when
we have equality in (\ref{eq:nproba}--c), and when all
$\Ppsi(\text{All }A\text{'s but one} \eq 1)$ are equal (which
gives the best possible value in (\ref{eq:nproba})).  We then would
have
\begin{equation}
  \eta\le\frac{n\Ppsi(\text{All }A\text{'s}
    \eq 1\text{ but }A_n)}%
  {(2n-1)\Ppsi(\text{All }A\text{'s}\eq 1\text{ but }A_n)}
  =\frac{n}{2n-1}.
\end{equation}
A lower bound cannot be obtained; we have proved the \emph{only if}
part. 

Again, states arbitrarily close to giving equality will be used in
the proof of the \emph{if} part. Given $\epsilon>0$, and using
$\theta=2\arctan(\epsilon)$, the quantum state (see \cite{PROOF}),
\begin{multline}
  \ket{\delta}=C\Big(\big(1-n\cos(\theta)\big)
  \ket{0_{B_1}\cdots0_{B_n}}\\
  +sin(\theta)\big(\ket{1_{B_1}0_{B_2}\cdots0_{B_n}}
  +\ldots
  +\ket{0_{B_1}\cdots0_{B_{n-1}}1_{B_n}}
  \big)\Big)
\end{multline}
(where $C$ is a normalization constant), and the rotation
\begin{equation}\label{eq:rot}
  \begin{pmatrix}
    \ket{0_{A_i}}\\\ket{1_{A_i}}
  \end{pmatrix}
  =
  \begin{bmatrix}
    \cos(\theta)&-\sin(\theta)\\\sin(\theta)&\cos(\theta)
  \end{bmatrix}
  \begin{pmatrix}
    \ket{0_{B_i}}\\\ket{1_{B_i}}
  \end{pmatrix}
\end{equation}
yields
\begin{align}
  \stepcounter{equation}
  & \Ppsi(\text{All }A\text{'s}\eq 1) =K\neq0
  \tag{\theequation a}\\
  & \Ppsi(\text{One }B
  \text{ and all other }A\text{'s}\eq 1)=K
  \tag{\theequation b}\\
  &\Ppsi(\text{Two or more }B\text{'s}\eq 1)= 0
  \tag{\theequation c}\\
  &\Ppsi(\text{All }A\text{'s
    but one} \eq 1)=K(1+\epsilon^2),\tag{\theequation d}
\end{align}
that is, (\ref{eq:nbound}) reduces to
\begin{equation}
  \eta\le\frac{nK(1+\epsilon^2)}{(2n-1)K}
  =\frac{n}{2n-1}(1+\epsilon^2).
\end{equation}
So when $\eta>\frac{n}{2n-1}$, there exists a quantum state that will
give a violation of the CH inequality.\pfbox
 
This $\ket{\delta}$ is in a similar fashion as in Theorem~2 not an
eigenstate, and a check in the three-site case shows that the the
violation from the best $\ket{\delta}$ is just below 70\% of that
obtained from the eigenstate. Unfortunately, the computational effort
required for these comparisons increases rapidly with the number of
sites.

To conclude, we have seen that in the two-site CH inequality, quantum
mechanics yields a violation only if $\eta>\frac{2}{3}$, and thus,
by the result in \cite{Pitowsky89}, the lowest bound we can obtain in
\emph{any} Bell inequality for two sites and two detector settings is
$\frac{2}{3}$. For $n$-site situations the picture is not so clear,
because of the lack of similar results.

It should be said that although the current paper yields low bounds on
efficiency, these bounds are only proven here for 100\% visibility.
While the inequalities are valid irrespective of visibility, extra
background events will reduce, and then prevent, a violation of the
inequality. The usable visibility depends on the absolute size of the
violation of the inequalities (i.e., $K$). In \cite{EBERHARD} it is
clear that lowering the bounds on $\eta$ puts higher demand on the
visibility. In general, the same behavior is to be expected in an
$n$-particle setup.

Nevertheless, the best bound possible to obtain from the $n$-particle
CH inequalities presented here is that when
\begin{equation}
  \eta>\frac{n}{2n-1},
\end{equation}
there is a quantum state yielding a violation in the corresponding
$n$-particle experiment. This enables us to state 

\emph{Theorem 5: (A bound on $\eta$)}
Whenever
\begin{equation}
  \eta>\frac{1}{2},
\end{equation}
there exists an $n$ and an $\epsilon$, so that the $n$-particle
quantum state $\ket{\delta}$ given above violates the corresponding
$n$-particle CH inequality.

Or in other words, there are experiments for which the
quantum-mechanical predictions violate local realism whenever the
efficiency exceeds $\frac{1}{2}$.

\section*{Acknowledgements}

J.S.\ would like to thank Craig Savage for useful discussions on the $n=3$
case.

\end{multicols}

\begin{thebibliography}{4}
\bibitem{EQUAL} If $\Ppsi(A_1 \eq 1)\neq\Ppsi(A_2 \eq
  1)$, equality in (\ref{eq:proba}--d) implies one of the terms in
  the numerator is less than the other, while two of the terms in the
  denominator are less than the third, yielding a higher bound.
\bibitem{PROOF} We want a symmetrical state $\ket{\delta}$ such that
  \begin{gather}
    \stepcounter{equation}
    \langle 1_{B_i}1_{B_j}\ket{\delta}
    =0,\quad i\neq j
    \label{eq:need0}\tag{\theequation a}\\
    \langle 1_{A_1}\cdots1_{A_{n-1}}1_{B_n}\ket{\delta}
    =\langle 1_{A_1}\cdots1_{A_n}\ket{\delta},
    \label{eq:need1}\tag{\theequation b}\\
    \langle 1_{A_1}\cdots1_{A_{n-1}}0_{A_n}\ket{\delta}
    =\epsilon\langle 1_{A_1}\cdots1_{A_n}\ket{\delta}.
    \label{eq:need2}\tag{\theequation c}
  \end{gather}
  Given (\ref{eq:need0}) and symmetry, we have
  \begin{multline}
    \ket{\delta}=a\ket{0_{B_1}\cdots0_{B_n}}\\
    +b\big(\ket{1_{B_1}0_{B_2}\cdots0_{B_n}}
    +\ldots
    +\ket{0_{B_1}\cdots0_{B_{n-1}}1_{B_n}}
    \big).
  \end{multline}
  Inserting this in (\ref{eq:need1}), using the general rotation in
  equation~(\ref{eq:rot}) and solving for $a$, we obtain
  \begin{equation}
    a=b\frac{1-n\cos(\theta)}{\sin(\theta)}.
  \end{equation}
  By use of (\ref{eq:need2}), we may solve for $\epsilon$:
  \begin{equation}
    \epsilon= \frac{\cos(\theta)-1}{\sin(\theta)}
      =\frac{2\sin^2(\frac{\theta}{2})}%
      {2\sin(\frac{\theta}{2})\cos(\frac{\theta}{2})}
      =\tan(\frac{\theta}{2}).
  \end{equation}
  We then have the probabilities we want from $\ket{\delta}$ and our
  rotation.
\end{thebibliography}
\end{document}